\documentclass[%
 reprint,
superscriptaddress,
 amsmath,amssymb,
 aps,
]{revtex4-2}

\usepackage{epsfig,dsfont,amssymb,amsmath,amsthm,amsfonts,amsbsy,mathrsfs}
\usepackage{graphicx}
\usepackage{color}
\usepackage{bm}
\usepackage{multirow}
\usepackage{natbib}

\usepackage{tikz}
\usetikzlibrary{arrows,shapes,chains}

\newcommand{\E}{{\mathbf e}}

\begin{document}

\title{Entropy production of run-and-tumble particles}

\author{Matteo Paoluzzi}
\address{Istituto per le Applicazioni del Calcolo del Consiglio Nazionale delle Ricerche, Via Pietro Castellino 111 80131 Napoli, Italy.}

\author{Andrea Puglisi}
\address{
Istituto dei Sistemi Complessi-Consiglio Nazionale delle Ricerche, and Dipartimento di Fisica,
Sapienza Universit\`a di Roma, Piazzale A. Moro 2, I-00185, Rome, Italy}

\author{Luca Angelani}
\address{
Istituto dei Sistemi Complessi-Consiglio Nazionale delle Ricerche, and Dipartimento di Fisica,
Sapienza Universit\`a di Roma, Piazzale A. Moro 2, I-00185, Rome, Italy}


\begin{abstract}
We analyze the entropy production in run-and-tumble models. 
After presenting the general formalism in the framework of the Fokker-Planck equations
in one space dimension, we derive 
some known exact results in simple physical situations (free run-and-tumble particles and 
harmonic confinement).
We then extend the calculation to the case of anisotropic motion (different speeds and tumbling rates
for right and left oriented particles), obtaining exact expressions of the entropy production rate.
We conclude by discussing the general case of heterogeneous run-and-tumble motion described by 
space-dependent parameters and extending the analysis to the case of $d$-dimensional motions.
\end{abstract}

\maketitle



\section{Introduction}

Active matter is a recently established research field in statistical physics~\cite{marchetti2013hydrodynamics}. It includes systems made of (typically) many particles endowed with self-propulsion, the most prominent examples coming from biology, e.g. microswimmers or motile cells at the micro-scale~\cite{elgeti2015physics} or birds and pedestrians at the macro-scale~\cite{cavagna2018physics}, but encompasses also motile artificial particles at all scales~\cite{callegari2023playing}.  Motility - which is a conversion of energy from some fuel/reservoir into motion of each particle, is a fascinating ingredient for theoretical physics, as it implies a source of time-reversal symmetry breaking in the bulk of the systems 
\cite{Angelani_2011,battle2016broken,gnesotto2018broken,maggi2023thermodynamic}, 
different from the usual forcing coming from the boundaries which occurs in older examples of out-of-equilibrium systems such as fluids under the action of externally imposed gradients (e.g. heat flow, convection, turbulence, etc.)~\cite{de2013non,livi2017nonequilibrium}. 

The interest of statistical physics for those systems is both at the level of a single active particle and at the level of large populations of active particles, since in both cases the lack of thermodynamic equilibrium triggers the appearance of unexpected phenomena 
\cite{ramaswamy2010mechanics,fodor2018statistical,Angelani2024EPJE}. 
A single self-propelling particle {\em hides} a complex arrangement of several internal degrees of freedom such as molecular motors actuating flagella, as in bacteria or sperms: it, therefore, may require non-trivial stochastic modeling, in contrast with passive Brownian particles~\cite{bechinger2016active}. A population of motile particles may exhibit collective behaviors that are not allowed when the motility ingredient is removed, typical examples being the polarisation transition of aligning active particles~\cite{toner2005hydrodynamics} and the motility-induced phase separation for purely repulsive active particles~\cite{PhysRevLett.100.218103,cates2015motility}. 

One of the questions concerning the non-equilibrium statistical physics of the single active particle is how to characterize the dissipation occurring because of the time-reversal symmetry breaking induced by the self-propulsion mechanism~\cite{o2022time}. A relevant approach to this problem is given by stochastic thermodynamics, which equips the theory of stochastic processes with a mesoscopic (fluctuating) definition of work, heat, and entropy production, including a fluctuating version of the second principle of thermodynamics~\cite{sekimoto2010stochastic,seifert2012stochastic,peliti2021stochastic}. 
The application of stochastic thermodynamics to single active particles has been developed in the recent years, starting from models with continuous noise~\cite{fodor2016far,marconi2017heat,shankar2018hidden,dabelow2019irreversibility,caprini2019entropy} such as Active Ornstein-Uhlenbeck Particles (AOUP) and Active Brownian Particles (ABP), and only more recently it has been addressed also for time-discontinuous models such as Run-and-Tumble particles (RT)~\cite{razin2020entropy,Cocco2020,Fry2022}. Such a model is considered a better description of certain biophysical systems, for instance, the  {\it E. coli} bacteria which has a re-orientation dynamics dominated by sudden changes rather than rotational diffusion \cite{Ecoli_Berg,RWinBio_Berg}. The less smooth mathematical structure of the model makes the problem interesting: for instance, ABP and AOUP have a finite entropy production even when traslational thermal diffusion - often considered negligible in real applications - is sent to zero in the model. On the contrary, a RT particle - under the influence of an external potential - in the limit of zero temperature becomes {\em strongly} time-irreversible, meaning that the time-reversed of an observable trajectory in general is not observable, corresponding to an infinite entropy production~\cite{cerino2015entropy}. 
The divergence is healed when a finite diffusivity $D>0$ is considered: typically - as seen also in this paper - the steady state entropy production diverges for $D \to 0$. Morally this corresponds to the fact that a model for active particles may have a finite rate for energy dissipation $\dot W$ even at zero temperature $T=0$ and therefore it is not a paradox to find a divergence for the entropy production rate, expected on general grounds to be $\dot W/T$. A closer look at the problem, however, suggests that in many cases - particularly in biology - all energy conversion processes occurring inside an active particle are triggered by thermal processes (e.g. the dynamics of motor proteins is fueled by ATP molecules but the energy barriers among the protein configurations cannot be overcome at $T=0$) and therefore one could expect $\dot W \sim T$ so that one might obtain
a finite entropy production rate 
in the limit $T \to 0$. This problem is however not the scope of this paper and the question will be addressed in future research.
The entropy production for run-and-tumble particles confined to move into a one-dimensional box has been the subject of~\cite{razin2020entropy}, following the recipe given in~\cite{tome2006entropy}. Here we revisit this problem with a more straightforward derivation.

The structure of the paper is the following. In Section II, we review the minimal ingredients for the definition of entropy production of Markov processes described by a Fokker-Planck equation. In Section III, we discuss entropy production for RT particles in 1D, starting with some known results re-derived more straightforwardly, i.e. free RT particles and then RT particles in a harmonic potential. In Section IV, we give the expression for anisotropic models, i.e. RT particles in 1D with different tumbling rates and/or different self-propulsion velocities in the two directions of motion. 
In Section V, we give a more general treatment which  includes several cases of practical interest
and in Section VI, we extend the calculation to the $d$-dimensional case.
Section VII is devoted to conclusions.  

\section{Theoretical set-up within the Fokker-Planck equation}
Here we briefly recall the theoretical 
framework for the computation of the entropy production rate in stochastic processes
governed by Fokker-Planck like equations~\cite{seifert2012stochastic}. 
Denoting with $S(t)$ the entropy of the system at the time $t$,
we can decompose the rate of change of the entropy into
two terms, $\Pi$ and $\Phi$, as
\begin{equation} \label{change}
\dot{S} = \frac{d S}{d t} = \Pi - \Phi     \, ,
\end{equation}
where $\Pi$ is the entropy production due to irreversible processes inside 
the system and $\Phi$ is the entropy flux from the system to the environment.
The entropy production $\Pi$ is non-negative while $\Phi$ can have either sign.

We consider a generic stochastic process describing a particle moving in a one-dimensional space.
The probability density function (PDF) $P(x,t)$ to find the particle at position $x$ at time $t$
obeys the following continuity equation
\begin{equation} \label{continuity}
\partial_t P(x,t) = -\partial_x J(x,t)     
\end{equation}
where $J(x,t)$ is the current and  $\partial_t$ and $\partial_x$ denote, respectively, time and space derivative.
In the case of the Fokker-Planck equation, one has the following
constitutive relation linking the current to the probability
\begin{equation} 
\label{FP}
J(x,t) =  [ \mu f(x) - D \partial_x ] P(x,t)   \, ,   
\end{equation}
with $D$ the diffusion constant, $f(x)$ a generic space-dependent mechanical force acting on the particle
and $\mu$ the particle mobility.

The Gibbs entropy $S(t)$ of the distribution $P(x,t)$ is defined as
\begin{equation}
S(t) = -\int dx \, P(x,t) \log P(x,t)     \, ,
\end{equation}
and the rate of the entropy change reads
\begin{align}
\dot{S}(t) &= -\int dx \, \dot{P}(x,t) \left[ 1 + \log P(x,t) \right] \nonumber \\ \nonumber
&= \int dx \, \partial_x J(x,t) \left[ 1 + \log P(x,t) \right] \\ 
& = - \int dx \, J(x,t) \partial_x \log P(x,t) \; , \label{entropy_change}
\end{align}
where we have used the
continuity equation (\ref{continuity})
and integration by parts assuming vanishing distributions at boundaries.
By using the relation (\ref{FP}), we can write
\begin{equation}
\frac{J(x,t)}{D P(x,t)} = \frac{\mu f(x)}{D} - \partial_x \log P(x,t)    
\end{equation}
and thus the expression for $\dot{S}(t)$ becomes
\begin{equation}
\dot{S} = - \int dx \, \frac{J(x,t)}{D} \left[ \mu f(x) - \frac{J(x,t)}{P(x,t)} \right]    \; .
\end{equation}
We finally obtain the following forms of the entropy rates defined in  (\ref{change})
\begin{align}
\dot{S}(t) &= \Pi(t) - \Phi(t) \\    
\Pi(t) &= \int dx \, \frac{J^2(x,t)}{D P(x,t) } \\
\Phi(t) &=\ \frac{\mu}{D}\int dx \, J(x,t) f(x) .
\label{Phi1}
\end{align}
As a functional of $J$, we immediately realize that $\Pi(t)$ is non-negative, 
being the integrand proportional to $J^2$ with positive coefficients, while $\Phi$ can be either negative or positive. $\Pi$ is the entropy production rate that can be also computed through the Kullback-Leibler divergence between the probability of a path of the system with respect to the time-reversal one.

In the stationary regime, we can compute the entropy production rate $\Pi$
by noting that the rate of entropy change $\dot{S}$ must be zero
\begin{align}
    \dot{S}_{st}= 0 =  \Pi_{st}  - \Phi_{st} \, ,
\end{align}
and thus we can compute $\Pi$ through the expression for $\Phi$ since they equals on stationary trajectories
\begin{align}
\label{EPRST}
    \Pi_{st} = \Phi_{st} \; .
\end{align}
When the Brownian particle reaches equilibrium, as, for example, in the presence of
a confining potential $V(x)$,
the entropy production rate is zero
\begin{equation}
\Pi_{st} = \frac{\mu}{D} \int dx \, f(x) \left[ \mu f(x) - D \partial_x \right] P_{eq}(x) = 0    \, ,
\end{equation}
as is immediately clear considering that
$f(x) = -\partial_x V(x)$ and 
$P_{eq}(x) \propto e^{-\mu V(x)/D}$.
Instead, in the case of a driven Brownian particle, we have a finite production entropy.
Indeed, in this case, the constant force produces a drift velocity $v=\mu f$,
thus resulting in
\begin{equation}
\label{EPR_DB}
\Pi_{st} = \frac{v^2}{D}    \, ,
\end{equation}
as obtained from (\ref{FP}), (\ref{Phi1}) and (\ref{EPRST}).

\section{Run-and-tumble motion}
We now calculate  the entropy production in the case of run-and-tumble motions
in the presence of thermal noise. 
We consider a particle that alternates sequences of {\it run} motion and {\it tumble} events:
it moves at constant speed $v$
in a given direction until it tumbles at rate $\alpha$, randomly choosing the new direction of motion
\cite{Schnitzer1993,weiss2002some}
In the one-dimensional case analyzed here, 
there are only two possible directions of motion, let say rigth and left
(in the last Section VI we will generalize the analysis to higher dimensions).
We assume that the particle is also subject to a thermal noise, described
by a diffusion coefficient $D$. We will first treat the case of a free particle and then the motion in
a confining harmonic potential. 
We  derive in a simple way the exact expressions of the entropy production rates,
without resorting to the explicit solutions of the kinetic equations of motion,
reproducing the exact results known in the literature
\cite{Cocco2020,razin2020entropy,Fry2022,GMP2021}.
Unlike the previous section, for the sake of simplicity, here and in the following we will omit in the
reported equations the 
explicit dependence on the $x$ and $t$ variables of the various quantities.

\subsection{Free run-and-tumble particles}
We first analyze the case of a free run-and-tumble particle.
We indicate with $R(x,t)$ the probability density function to find the particle at position $x$ at the time $t$ moving towards the right, and with $L(x,t)$ the probability density function for the particle moving towards the left.
The coupled kinetic equations describing the run-and-tumble motion in the presence of thermal noise are 
\begin{align} \label{eq:rt_free}
    \partial_t R &= D \partial_x^2 R - v\partial_x R + \frac{\alpha}{2} \left( L - R \right)\\ 
    \partial_t L &= D \partial_x^2 L + v\partial_x L - \frac{\alpha}{2} \left( L - R \right) \; .
\end{align}
Once we introduce the currents
\begin{align}
    J_R &= v R - D \partial_x R \\ 
    J_L &= -v L - D \partial_x L \\
    J_{LR} &= \frac{\alpha}{2} \left( R - L\right)
\end{align}
we can write the equations of motion as follows
\begin{align} \label{eq:tr_free_2}
    \partial_t R &= - \partial_x J_R - J_{LR} \\ 
    \partial_t L &= - \partial_x J_L + J_{LR} \; .
\end{align}
The entropy $S$ is given by the sum of the two entropies
\begin{align}
\label{S1dsum}
    S &= S_R + S_L \\
    S_R &= -\int dx \, R \log R \\ 
    S_L &= -\int dx \, L \log L \; .
\end{align}
Once we performed the time derivative 
\begin{align}
    \dot{S}   &= \dot{S}_R + \dot{S}_L \\
    \dot{S}_R &= - \int dx \, \partial_t R \left( 1 + \log R \right)\\ 
    \dot{S}_L &=  - \int dx \, \partial_t L \left( 1 + \log L \right) \; .
\end{align}
Once we plug Eqs. (\ref{eq:tr_free_2}) we obtain
\begin{align}
    \dot{S}_R &= \int dx \, \left( \partial_x J_R + J_{LR}\right) \left( 1 + \log R\right)  \\ \nonumber 
   &=  - \int dx \, \frac{J_R}{R} \partial_x R + \int dx \, J_{LR} \left( 1 + \log R \right) \; .
\end{align}
and similarly 
\begin{align}
    \dot{S}_L &=  \int dx \, \left( \partial_x J_L - J_{LR}\right) \left( 1 + \log L\right)  \\ \nonumber 
    &=- \int dx \, \frac{J_L}{L} \partial_x L - \int dx \, J_{LR} \left( 1 + \log L \right) \; ,
\end{align}
having considered that distributions vanish at infinity.
Using the expressions for $J_{R,L}$, we can write
\begin{align}
    \frac{\partial_x R}{R}  &= \frac{1}{D} \left( v -\frac{J_R}{R}\right) \\ \nonumber 
    \frac{\partial_x L}{L} &= -\frac{1}{D} \left( v +\frac{J_L}{L} \right)
\end{align}
so that, upon neglecting boundary terms, we obtain
\begin{align}
    \dot{S} &=       \Pi - \Phi \\ \nonumber 
    \Pi     &=  \frac{1}{D} \int dx \, \left( \frac{J_R^2}{R} + \frac{J_L^2}{L}\right) + \frac{\alpha}{2} \int dx \, \left( R - L \right) \log \frac{R}{L} \\ \nonumber 
    \Phi    &= \frac{v}{D} \int dx \, \left( J_R -J_L \right) \; .
\end{align}
At the steady-state we get $\dot{S}=0$ so that 
$\Pi_{st}=\Phi_{st}$ and thus the entropy production rate is given by
\begin{align} \label{flux1}
    \Pi_{st} = \frac{v}{D} \int dx \, ( J_R - J_L )  \; .
\end{align}
Once we introduce
\begin{align}
\label{PRL}
    P      \equiv R + L \, ,\\
    Q \equiv R - L \, ,
\label{QRL}
\end{align}
with $\int dx P(x) = 1$, we obtain
\begin{align}
    J_R - J_L = v P - D \partial_x Q \, ,
\end{align}
so that
\begin{align}
\label{EPR_free}
    \Pi_{st} =\frac{v^2}{D} \; .
\end{align}
We note that the above result is the same obtained for a driven Brownian particle.
Indeed, we observe that a free run-and-tumble particle with diffusion can be viewed as a drift-diffusive 
particle going constantly in the direction parallel to its own driving force, even if such a force (proportional to the velocity of the particle) changes direction at random times. The process of tumbling is instantaneous and therefore does not add any contribution to the entropy production.

\subsection{Run-and-Tumble particles in harmonic potential}
We now consider the case of a run-an-tumble particle in a confining (harmonic) potential
\begin{equation}
    V(x) = \frac{k}{2} x^2 ,
\end{equation}
where $k$ is the potential stiffness. The Fokker-Planck equations for $R$ and $L$ are
\begin{align} \label{eq:harmR}
    \partial_t R &= - \partial_x J_R - J_{LR} \\ 
    \partial_t L &= - \partial_x J_L + J_{LR} \; ,
    \label{eq:harmL}
\end{align}
where 
\begin{align}
    J_R &= (v + \mu f  - D \partial_x) R \\ 
    J_L &= (-v +\mu f - D \partial_x) L \\
    J_{LR} &= \frac{\alpha}{2} \left( R - L\right)
    \label{JLR0}
\end{align}
and  $f(x)=-\partial_x V(x)=-kx$ is the force field .
Proceeding as before, we can write the entropy rate as
\begin{align}
    \dot{S} &=       \Pi - \Phi \\ \nonumber 
    \Pi     &=  \frac{1}{D} \int dx \, \left( \frac{J_R^2}{R} + \frac{J_L^2}{L}\right) + \frac{\alpha}{2} \int dx \, \left( R - L \right) \log \frac{R}{L} \\ \nonumber 
    \Phi    &= \frac{v}{D} \int dx \, \left( J_R -J_L \right) 
    - \frac{\mu k}{D} \int dx \, x \left( J_R +J_L \right) 
    \; .
\end{align}
In the steady state we have $\Pi_{st}=\Phi_{st}$ and, considering that $J=J_R+J_L=0$, we obtain
\begin{equation}
\label{EPR_v1}
    \Pi_{st} = \frac{v}{D} \int dx \, \left( J_R -J_L \right)
\end{equation}
By noting that 
\begin{equation}
J_R-J_L= vP-\mu kxQ -D\partial_x Q    
\end{equation}
where $P$ and $Q$ are defined in (\ref{PRL})-(\ref{QRL}), we have (considering  the normalization condition and the vanishing of the distributions 
at infinity)
\begin{equation}
\label{EPR_v2}
    \Pi_{st} = \frac{v}{D} \left( v + I \right) 
\end{equation}
where 
\begin{equation}
    I \equiv  -\mu k \int dx \  x Q
\end{equation}
From (\ref{eq:harmR}), (\ref{eq:harmL}) and (\ref{JLR0}), in the stationary regime we have
\begin{equation}
    \partial_x \left( J_R-J_L\right) = -\alpha Q \ ,
\end{equation}
and, multiplying by the force and integrating over  space, gives
\begin{equation}
\mu k \int dx \, x \partial_x \left( J_R-J_L \right) = \alpha I \ .
\end{equation}
Integrating by parts 
 we obtain
\begin{equation}
    \alpha I = - \mu k \int dx \left( J_R-J_L\right) = -\mu k \left( v + I \right)
\end{equation}
and then
\begin{equation}
    I = - \frac{\mu k v}{\alpha+\mu k}
\end{equation}
Substituting in (\ref{EPR_v2}) we finally obtain  the expression of the entropy production rate
\begin{equation}
\label{EPR_harm}
    \Pi_{st} = \frac{v^2}{D} \, \frac{\alpha}{\alpha +\mu k}
\end{equation}
The above expression is in agreement with that reported in \cite{Fry2022} 
-- see eq. (55) --
and also in \cite{GMP2021}, eq. (41), where it has been obtained using a path integral approach.  
We note that for $k=0$ we recover the previous expression (\ref{EPR_free}) valid for a free run-and-tumble particle.
It is remarkable that the above result has been obtained without resorting to the exact 
stationary solution of the run-and-tumble equations, 
which indeed in this case cannot be written in closed form \cite{Frydel2022}.

\section{Anisotropic run-and-tumble motion}
We extend here the analysis of the previous section to the  case of particles performing 
anisotropic run-and-tumble motion, i.e., 
we consider tumbling rates and speeds which depend on the orientation of the particle,
$\alpha_R \neq \alpha_L$ and  $v_R \neq v_L$.
These parameters are assumed to be constant in time and space, which will allow us to obtain 
exact results for the entropy production. 
In the next section we will relax the spatial homogeneity condition,
allowing the speeds and tumbling rates to depend explicitly on the variable $x$.
We treat here the case of motion in the presence of a harmonic potential 
$V(x) = \frac{k}{2} x^2$,
the free-case being recovered in the limit of null spring constant, $k=0$.
The Fokker-Planck equations for $R$ and $L$ are 
\begin{align} \label{FPR}
    \partial_t R &= - \partial_x J_R - J_{LR} \\ 
    \partial_t L &= - \partial_x J_L + J_{LR} \; ,
    \label{FPL}
\end{align}
where 
\begin{align}
\label{JR}
    J_R &= (v_R + \mu f  - D \partial_x) R \\ 
    \label{JL}
    J_L &= (-v_L +\mu f - D \partial_x) L \\
    \label{JLR}
    J_{LR} &= \frac{1}{2} \left(\alpha_R R - \alpha_L L\right)
\end{align}
and  $f(x)=-\partial_x V(x)=-kx$ is the force field. 
The entropy rate is 
\begin{align}
    \dot{S} &=       \Pi - \Phi \\ \nonumber 
    \Pi     &=  \frac{1}{D} \int dx \, \left( \frac{J_R^2}{R} + \frac{J_L^2}{L}\right) + \frac{1}{2} \int dx \, \left( \alpha_R R - \alpha _L L \right) \log \frac{R}{L} \\ \nonumber 
    \Phi    &= \frac{1}{D} \int dx \, \left( v_R J_R - v_L J_L \right) 
    - \frac{\mu k}{D} \int dx \, x \left( J_R +J_L \right) 
    \; .
\end{align}
In the steady state we have
\begin{equation}
\label{PI1}
    \Pi_{st} = \frac{1}{D} \int dx \, \left(v_R J_R - v_L J_L \right) .
\end{equation}
By using (\ref{JR}) and (\ref{JL}) we have
\begin{eqnarray}
\label{PI2}
    D \Pi_{st} &=& v_R^2 \int dx R  + v_L^2 \int dx L \nonumber \\
    &+& \mu k v_L \int dx x L 
    - \mu k v_R \int dx x R 
\end{eqnarray}
We now observe that the first two integrals in (\ref{PI2}) are given by
\begin{align}
\label{Req1}
    \int dx R & = \frac{\alpha_L}{\alpha_R+\alpha_L} \\
    \int dx L & = \frac{\alpha_R}{\alpha_R+\alpha_L} 
\label{Leq1}
\end{align}
as obtained considering the normalization condition of $P=R+L$ and that
the integral of $J_{LR}$ (\ref{JLR}) must be zero, as it follows from Fokker-Planck equations
in the stationary regime).\\

Now we consider the case $k>0$. From (\ref{FPR}) and (\ref{FPL}) in the stationary regime we have
\begin{equation}
    \partial_x \left(v_R J_R- v_L J_L\right) = -\frac{v_R+v_L}{2} 
    \left( \alpha_R R -\alpha_L L\right)
\end{equation}
and then, multiplying by $\mu k x$ and integrating over $x$
\begin{equation}
\mu k \int dx \ x \partial_x \left( v_R J_R- v_L J_L \right) = 
\frac{v_R+v_L}{2} \left( \alpha_R I - \alpha_L Y  \right) ,
\end{equation}
where 
\begin{align}
        I & \equiv  -\mu k \int dx \ x R \\
        Y & \equiv  -\mu k \int dx \ x L .
\end{align}
Integrating by parts 
we obtain
\begin{equation}
\label{IY2}
\mu k \int dx \, \left(v_R J_R - v_L J_L \right) = 
-\frac{v_R+v_L}{2} \left( \alpha_R I - \alpha_L Y  \right) \, .
\end{equation}
The quantities $I$ and $Y$ are related to each other.
Indeed, in the steady state, the total current is zero and then, 
using (\ref{JR}) and (\ref{JL}), we have
\begin{equation}
 0 = \int dx (J_R+J_L) = v_R \int dx R - v_L \int dx L + I + Y
\end{equation}
Using (\ref{Req1}) and (\ref{Leq1}) we get
\begin{equation}
\label{IY3}
    I+Y = \frac{v_L \alpha_R -v_R \alpha_L}{\alpha_R+\alpha_L}.
\end{equation}
Combining equations   (\ref{IY2}) and (\ref{IY3}) 
-- together with (\ref{Req1}) and (\ref{Leq1}) -- we obtain an equation 
for $I$, whose solution is
\begin{equation}
    I = \frac{\alpha_L}{\alpha_R+\alpha_L} \ 
    \frac{\alpha_R v_L -\alpha_L v_R -2 \mu k v_R}{2\mu k+\alpha_R+\alpha_L} .
\end{equation}
Using (\ref{IY3}) we obtain for $Y$
\begin{equation}
    Y = \frac{\alpha_R}{\alpha_R+\alpha_L} \ 
    \frac{\alpha_R v_L -\alpha_L v_R +2 \mu k v_L}{2\mu k+\alpha_R+\alpha_L} .
\end{equation}
Substituting in (\ref{IY2}) and using (\ref{PI1}), we finally arrive at the expression of the entropy production rate  for $k>0$:
\begin{equation}
\label{EPR_ART}
    \Pi_{st} = \frac{(v_R+v_L)^2}{D} \, \frac{\alpha_R \alpha_L}{(2\mu k+\alpha_R+\alpha_L)(\alpha_R+\alpha_L)} .
\end{equation}
Defining the average speed $v=(v_R+v_L)/2$, the average tumbling rate 
$\alpha=(\alpha_R+\alpha_L)/2$ and the tumbling rate semidifference $\delta=(\alpha_R-\alpha_L)/2$,
the EPR takes the simple form
\begin{equation}
\label{EPR_ART2}
    \Pi_{st} = \frac{v^2}{D} \, \frac{\alpha^2 -\delta^2}{\alpha (\mu k+\alpha)}\  .
\end{equation}
\begin{figure}[t!]
\includegraphics[width=.8\linewidth] {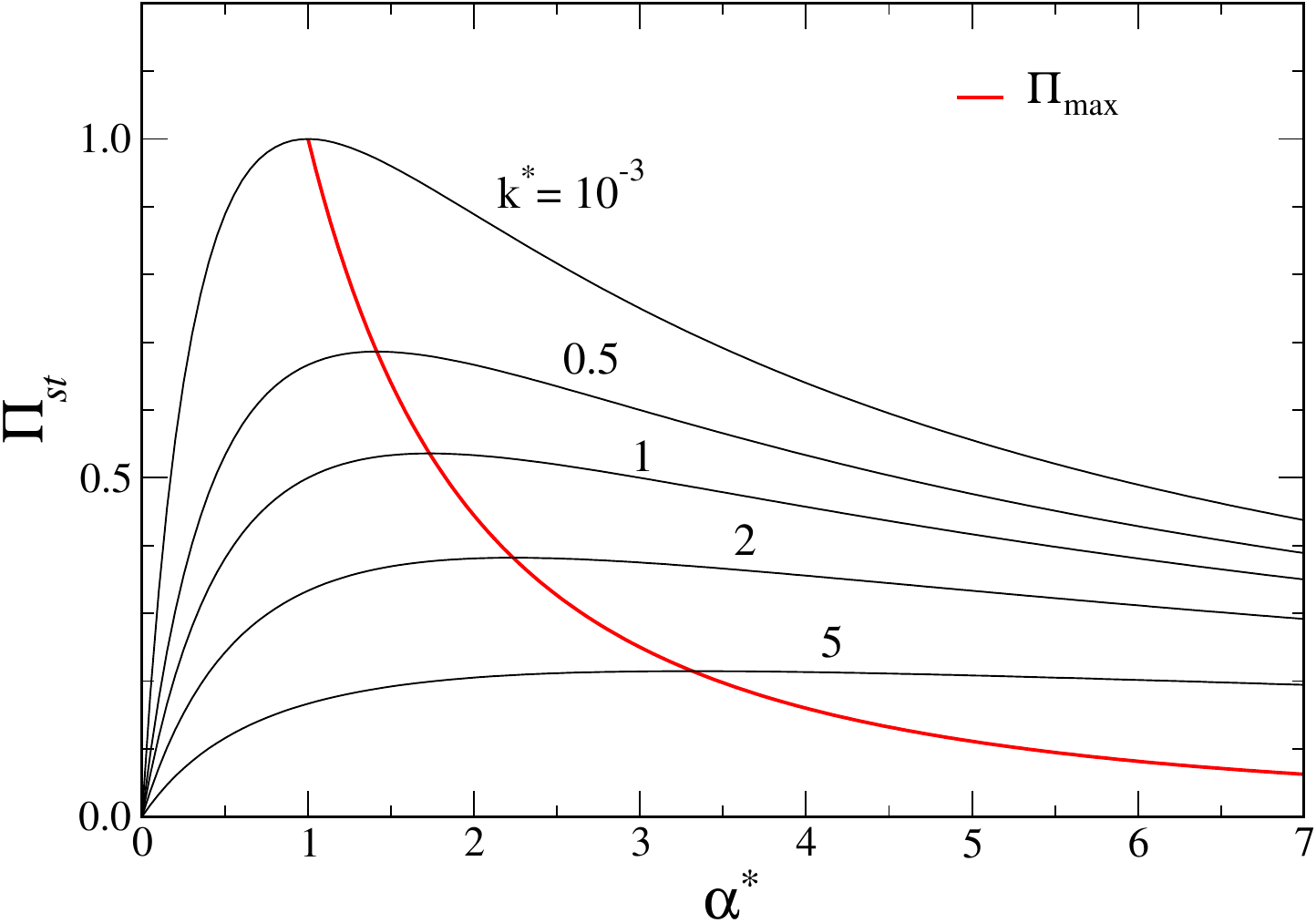}
\caption{\label{fig1}
Entropy production rate at stationarity -- see (\ref{EPR_ART}) in the text -- as a function of the relative
tumbling rate $\alpha^*=\alpha_R/\alpha_L$ for different values of the reduced harmonic constant 
$k^*=k/\alpha_L=10^{-3}, 0.5, 1, 2, 5$.
The red line is the maximum EPR $\Pi_{max}$  vs. $\alpha^*_{max}(k^*)$.
Units are such that $v=1$, $D=1$, $\mu=1$.
}
\end{figure}
For $\alpha_R=\alpha_L$, i.e., $\delta=0$, the EPR reads
\begin{equation}
    \Pi_{st} = \frac{v^2}{D} \, \frac{\alpha}{\mu k+\alpha} , \qquad \delta=0\ ,
\end{equation}
similar to the expression obtained in the isotropic case (\ref{EPR_harm})
with the average speed  $v=(v_R+v_L)/2$.\\

In the free case, the EPR can be computed directly by putting $k=0$ into Eq.~\eqref{PI2}, that - together with Eqs.~\eqref{Req1}-\eqref{Leq1} - leads to
\begin{equation} \label{EPR_ART_FREE}
    \Pi_{st}=\frac{1}{D}\frac{\alpha_L v_R^2+\alpha_R v_L^2}{\alpha_R+\alpha_L}.
\end{equation}
We first note that the limit $k\to 0$ of Eq.~\eqref{EPR_ART} is different from~\eqref{EPR_ART_FREE}, i.e. it is singular. This has already been noticed, in the case $v_R=v_L$, in~\cite{bao2023improving}.
The reason is that, in the free anisotropic case, a residual total current is present even in the steady state (i.e. asymptotically in time) and that is an additional source of dissipation. Such a current vanishes as soon as $k>0$, even very small. Formula~\eqref{EPR_ART_FREE} gives for $\alpha_R=\alpha_L=\alpha$:
\begin{equation}
      \Pi_{st}=\frac{1}{D}\frac{v_R^2+v_L^2}{2},
\end{equation}
    a result already obtained in~\cite{Cocco2020} by means of trajectory-based approach.
When $v_R\!=\!v_L\!=\!v$ we instead obtain~\cite{bao2023improving}
\begin{equation}
    \Pi_{st}=\frac{v^2}{D},
\end{equation}
i.e. the same result for the free isotropic case, remarkably independent from the tumbling rates.

It is worth noting that, in the general case, for fixed external potential
($k>0$) the EPR reaches its maximum value $v^2/D$
in the symmetric case ($\delta=0$) and for large tumbling rate ($\alpha \to \infty$).
However, some interesting behaviors of the EPR are obtained by considering some 
parameters fixed. 
While it is true that, fixing $k$ and $\alpha$, the maximum EPR 
$v^2(1+\mu k/\alpha)^{-1}/D$
is always obtained for $\delta=0$, in the case of fixed $k$ and $\alpha_L$ one has 
that the maximum EPR is reached for $\alpha^*=\alpha_R/\alpha_L>1$
(see Figure \ref{fig1}). 
The same would happen by fixing the value of $\alpha_R$, with the relative tumbling rate
given by $\alpha^*=\alpha_L/\alpha_R$.

\section{General run-and-tumble motion}
Let us now treat the very general case of anisotropic and heterogeneous run-and-tumble motion.
We consider the possibility that, not only tumbling rates and
speeds could be different for left and right oriented particles,
but they could also depend on the spatial variable, 
$\alpha_{R,L}(x)$ and $v_{R,L}(x)$.
Moreover, we consider the presence of a generic external force $f(x)$, 
not necessarily originated by a confining quadratic potential.
In this general case the Fokker-Planck equations for $R$ and $L$ can be written as
(for the sake of simplicity we omit the dependence on $x$-variable of the physical  parameters)
\begin{align} \label{FPRgen}
    \partial_t R &= - \partial_x J_R - J_{LR} \; , \\ 
    \partial_t L &= - \partial_x J_L + J_{LR} \; ,
    \label{FPLgen}
\end{align}
where 
\begin{align}
\label{JRgen}
    J_R &= (v_R + \mu f  - D \partial_x) R \; ,\\ 
    \label{JLgen}
    J_L &= (-v_L +\mu f - D \partial_x) L \; ,\\
    \label{JLRgen}
    J_{LR} &= \frac{1}{2} \left(\alpha_R R - \alpha_L L\right) \; .
\end{align}
The entropy rate is 
\begin{align}
\label{Sgen1d}
    \dot{S} &=       \Pi - \Phi \\ \nonumber 
    \Pi     &=  \frac{1}{D} \int dx \, \left( \frac{J_R^2}{R} + \frac{J_L^2}{L}\right) + \frac{1}{2} \int dx \, \left( \alpha_R R - \alpha _L L \right) \log \frac{R}{L} \\ \nonumber 
    \Phi    &= \frac{1}{D} \int dx \, \left( v_R J_R - v_L J_L \right) 
    + \frac{\mu}{D} \int dx \, f \left( J_R +J_L \right) 
    \; .
\end{align}
In the steady state we have
\begin{equation}
\label{PI1gen}
    \Pi_{st} = \frac{1}{D} \int dx \, \left(v_R J_R - v_L J_L \right) 
        + \frac{\mu}{D} \int dx \, f \left( J_R +J_L \right) \: .
\end{equation}
In the case of vanishing flows at steady-state  $J_R+J_L=0$ 
(as occurs, for example, in the presence of confining potentials) 
the above expression is formally identical to the one 
obtained in the previous section (\ref{PI1}),
but now the parameters $v_{R,L}$ and  $\alpha_{R,L}$ 
are explicitly space-dependent quantities.
In the general case it is not possible to obtain exact expressions of the EPR
and we need to resort to numerical solution of kinetic equations or numerical simulations
of the trajectories of the run-and-tumble particles.\\
We conclude this section by mentioning some particular case studies, 
that are interesting for their physical or biological relevance.

\noindent
{\bf Photokinetic bacteria.}
Photokinetic bacteria are characterized by spatially varying speed which depends on
 local light intensity $I$ \cite{Elife2018}. 
For static non-homogeneous light fields $I(x)$ we can
describe the particle dynamics through a space dependent speed $v(x)$ \cite{AngGar2019}
(we assume equal left and right speeds)
\begin{equation}
    v(x) = v (I(x)) \, .
\end{equation}

\noindent
{\bf Chemotaxis.} 
In the presence of nutrient concentration some motile bacteria modify their tumble rates 
to effectively direct their movement toward the food source
\cite{Ecoli_Berg,Schnitzer1993}.
We can describe such a phenomenon by expressing the tumble rates in terms 
of the chemotactic field $c(x)$. In the limit of weak concentration gradient we
can write \cite{Schnitzer1993,Cates_2012,FP_Ang2014}
\begin{eqnarray}
    \alpha_R(x) &=& \alpha - \gamma v \partial_x c(x) , \\ 
    \alpha_L(x) &=& \alpha + \gamma v \partial_x c(x) , 
\end{eqnarray}
with $\gamma$ measuring the strength of particle reaction to chemical gradients
and we have assumed equal speeds $v_R=v_L=v$.

\noindent
{\bf Generic confining potentials.} 
In the previous sections we have analyzed the case of a force field $f(x)=-\partial_x V(x)$ 
originated by  quadratic potentials $V(x) \propto x^2$.
It would be interesting to consider generic confining potentials
\cite{Dhar2019,Guéneau_2024}
\begin{equation}
    V(x) = a |x|^p \, , \qquad p\geq 1 \, ,
\end{equation}
and investigate the dependence on the exponent $p$.
Also of interest is the case of double-well potentials
\begin{equation}
    V(x) = a x^4 -b x^2 +c x \, ,
\end{equation}
in its symmetric ($c=0$) or asymmetric ($c\neq0$) version. 

\noindent
{\bf Ratchet potentials.} 
Finally, we mention the study of ratchet effect \cite{Angelani_2011}.
In this case, the active motion
takes place in the presence of a periodic asymmetric potential, giving rise
to unidirectional motion with a stationary flow of particles, $J_R+J_L\neq0$.
In the case of a piecewise-linear ratchet potential,
the entropy production for particles with equal tumbling rates and speeds has been analyzed in \cite{RZ2023}.

\section{Run-and-tumble motion in $\mathbb{R}^d$}
So far we have considered the case of one-dimensional motions.
Here we extend the analysis to $d$-dimensional run-and-tumble walks.
We consider a particle that, in the free case, moves along straight lines with velocity ${\bf v}= v \E$, 
where $v$ is the speed and $\E$ a unit vector in $\mathbb{R}^d$, 
and changes its direction of motion $\E$ with rate $\alpha$. 
We will first derive the general expression of the EPR considering generic space- and orientation-dependent speed and tumbling rate,
$v({\bf x},\E)$ and $\alpha({\bf x},\E)$. Then we will specialize to the simple case of 
constant $v$ and $\alpha$, showing the exact expression of the EPR in the presence of a harmonic potential.\\
By denoting with $p({\bf x},t;\E)$  the PDF to find the particle 
at position ${\bf x}\in\mathbb R^d$ at time $t$ with velocity orientation $\E$,  
the kinetic equation 
of the run-and-tumble motion can be written as \cite{martens2012probability} 
\begin{equation}
\partial_t p= - \nabla \cdot {\bf j} 
+\alpha (\mathbb{P} - 1) p \ ,
\label{Pkind}
\end{equation}
where the current ${\bf j}$ is 
(we consider the presence of thermal noise and generic force field $f({\bf x)}$)
\begin{equation}
    {\bf j} = \left( -D \nabla + v \E +\mu f \right) p \ ,
\label{Jdim}
\end{equation}
and we have introduced the projector operator 
\begin{equation}
    \mathbb{P} f({\bf x},t;\E) = \int \frac{d\E}{\Omega_d} \ f({\bf x},t;\E) ,
\end{equation}
with $\Omega_d\!=\!2 \pi^{d/2}/\Gamma(d/2)$ the solid angle in $d$-dimension.
Hereafter we 
consider normalization condition $\int d{\bf x} \ d\E \ p({\bf x},t;\E) =1$.
We define the total entropy $S$ as -- generalizing \eqref{S1dsum}
\begin{equation}
    S(t) = \int d\E \ s(t;\E) \ ,
\end{equation}
where the orientation dependent entropy $s$ is
\begin{equation}
    s(t;\E) = - \int d{\bf x} \ p({\bf x},t;\E) \log p({\bf x},t;\E) .
\end{equation}
By performing a derivation similar to that of the previous section we arrive at the 
expression of the entropy rate 
\begin{align}
    \dot{S} &=       \Pi - \Phi \ , \\ \nonumber 
    \Pi     &=   \int d{\bf x} \int d\E \ \left[ \frac{|{\bf j}|^2}{Dp} + 
     \alpha(1-{\mathbb P}) p \log p \right] \ , \\ \nonumber 
    \Phi    &=   \frac{1}{D} \int d{\bf x} \int d\E \ \left( v \E +
    \mu {\bf f}\right)  \cdot {\bf j}  \; ,
\end{align}
which generalize to dimension $d>1$ the expressions previously obtained
(\ref{Sgen1d}).
In the steady state we have $\Pi=\Phi$, and, assuming a null net current 
$\int d\E\ {\bf j}=0$, we have
that the EPR reads
\begin{equation}
    \Pi_{st} = \frac{1}{D} \int d{\bf x} \int d\E \ v \E \cdot {\bf j} \ .
\label{Pidim}
 \end{equation}
The results obtained so far are valid in the general non-homogeneous and non-isotropic case,
i.e., for generic $v({\bf x},\E)$ and $\alpha({\bf x},\E)$. 
We now specify the calculation to the case of constant parameters $v$ and $\alpha$,
extending the analysis of planar motions in \cite{Fry2022}
to ${\mathbb R}^d$ with generic $d>1$.
By using (\ref{Jdim}) we can write the EPR as
\begin{equation}
\Pi_{st} = \frac{v^2}{D} \left[ 1+ \frac{\mu}{v} \int d{\bf x}\int d\E \ p \ \E \cdot {\bf f} 
\right] \ ,
\label{Pidim2}
\end{equation}
having used the normalization condition and neglecting boundary terms.
Consider below a force field due to a harmonic potential, i.e., ${\bf f}=-k {\bf x}$. 
By substituting (\ref{Jdim}) in (\ref{Pkind}) in the stationary regime,
multiplying by 
$\E \cdot {\bf x}$, integrating over $d{\bf x}$ and $d\E$ and using integration by parts, we 
arrive at an equation for the quantity 
\begin{equation}
    I \equiv \int d{\bf x} \int d\E \ p \ \E \cdot {\bf x} \ ,
\end{equation}
appearing in the second term of (\ref{Pidim2}), which is
\begin{equation}
    (\alpha-dk\mu)I=v-k\mu(1+d) I\ ,
\end{equation}
leading to 
\begin{equation}
    I = \frac{v}{\alpha +\mu k} \ .
\end{equation}
Substituting in (\ref{Pidim2}) we finally obtain the expression of the EPR 
\begin{equation}
    \Pi_{st} = \frac{v^2}{D} \frac{\alpha}{\alpha + \mu k} ,
\end{equation}
which is the same as that obtained in the one-dimensional case (\ref{EPR_harm})
and is therefore independent of spatial dimensions.

\section{Conclusions}

We have computed the average entropy production rate, in the steady state, for a non-interacting run-and-tumble particle in several different physical setups. The general strategy is to start from the kinetic equations and then compute the entropy flux, identical to the entropy production in a steady state. The entropy flux - in the absence of a total net current (e.g. in confined or spatially symmetric situations) - is seen to be proportional to the difference of left-right currents $J_L,J_R$, weighted by the left-right speeds $v_L,v_R$ (Eqs.~\eqref{flux1},~\eqref{EPR_v1},~\eqref{PI1} in the different situations). The left-right currents endow also a dependence upon the tumbling rates. Such a weighted difference can be computed, in most of the considered situations, {\em without} computing the single currents but going directly to compute their weighted difference. 
This is a shortcut which allows us to revisit 
the free and harmonically confined cases, which already had a solution in the literature. 
The power of the method enables us to compute the entropy production rate also in non-symmetric setups where the tumbling rates and the velocities are different when particles go to the left or to the right. A discussion of the more general case where all parameters are space-dependent has also been presented, but explicit results cannot be usually obtained: a few cases of physical relevance are discussed with some detail. 
We have finally extended the calculation to the case of run-and-tumble motions in a $d$-dimensional space,
showing the formal expression of the EPR in the general case of space- and orientation-dependent parameters 
and reporting the exact solution in the case of harmonic potential and 
constant speed and tumbling rate.
Future research should focus on the entropy production for interacting RT systems 
exhibiting Motility-Induced Phase Separation \cite{cates2015motility}, 
where non-equilibrium density fluctuations have been investigated usually starting 
from opportune coarse-graining descriptions \cite{PhysRevX.7.021007,PhysRevLett.124.240604,PhysRevE.105.044139}. 
Finally, the theoretical framework considered here might be tested against experiments such as the ones recently done on different biological systems where EPR can be computed in a model-independent fashion \cite{di2024variance,PhysRevLett.129.220601}.

\begin{acknowledgments}
LA acknowledges financial support from the Italian Ministry of University and Research (MUR) 
under  PRIN2020 Grant No. 2020PFCXPE.
MP acknowledges NextGeneration EU (CUP B63C22000730005), Project IR0000029 - Humanities and Cultural Heritage Italian Open Science Cloud (H2IOSC) - M4, C2, Action 3.1.1.
\end{acknowledgments}

%


\end{document}